# Defect-Mediated Nucleation and Dynamics across the Phase Transition in the Excitonic Insulator Candidate $Ta_2NiSe_5$

Guilherme Rodrigues-Fontenele[1], Gabriel Fontenele[1], Ângelo Malachias[1], Rogério Magalhães-Paniago[1,*]

1 Physics Department, Federal University of Minas Gerais (UFMG), Belo Horizonte, Minas Gerais, Brazil.

* Corresponding author: rogerio.paniago0@gmail.com

**ABSTRACT:** $Ta_2NiSe_5$ is a quasi-one-dimensional material that exhibits a structural and electronic phase transition from a low-temperature monoclinic (semiconductor) to a high-temperature orthorhombic (semimetal) phase at $T_C \approx 326$ K. Here, we used variable-temperature scanning tunneling microscopy and spectroscopy to resolve the phase transition spatially, identifying the distinct spectroscopic signatures of the monoclinic and orthorhombic phases in pristine regions and near isolated point-defects and step edges. Although the phase transition of $Ta_2NiSe_5$ is generally regarded as second-order, it has previously been described as exhibiting martensitic-like characteristics. This implies that the transformation may proceed via spatial phase coexistence and domain boundaries rather than through a continuous evolution. Our surface-sensitive measurements confirm this scenario, retrieving the coexistence and evolution of monoclinic and orthorhombic domains in real space. Upon heating through $T_C$, we find that the two phases coexist as spatially segregated regions over extended timescales, separated by well-defined boundaries that evolve via localized nucleation and growth, rather than a spatially uniform transformation. In pristine regions, the orthorhombic phase nucleates anisotropically, perpendicular to the Ta-Ni-Ta chains, whereas point-defects and step edges act as local nucleation centers that promote the transition and suppress this intrinsic anisotropy. These results provide direct real-space visualization of how surface-specific structural and electronic variations, together with local disorder, modify the martensitic-like phase transition in $Ta_2NiSe_5$.

## INTRODUCTION

Phase transitions in low-dimensional materials continue to attract significant interest, giving rise to novel electronic phenomena including metal-insulator transitions, charge density waves, and superconductivity.[1–5] Additionally, these transitions often involve competing polymorphs separated by only small free-energy differences, making their stability highly sensitive to local perturbations such as strain and defects.[6,7] In this context, $Ta_2NiSe_5$ is a quasi-one-dimensional (1D) material that undergoes a coupled structural and electronic phase transition from a low-temperature monoclinic (semiconductor) to a high-temperature orthorhombic (semimetal) structure at $T_C \approx 326$ K.[8–10] This compound has been widely discussed as an excitonic insulator candidate – a correlated phase arising from the spontaneous condensation of excitons in narrow-gap semiconductors or semimetals – with its monoclinic structure often associated with this putative phase.[11–14] In addition to its fundamental correlated-electron physics, $Ta_2NiSe_5$ has also attracted attention in optoelectronic applications, particularly in devices that benefit from its narrow bandgap of ~ 0.20 eV and intrinsic structural anisotropy, such as polarization-sensitive photodetectors and phototransistors.[15–21]

The phase transition in $Ta_2NiSe_5$ is generally classified as second-order. Early support for a continuous transition came from Di Salvo et al., who reported reversible anomalies in electrical resistivity and magnetic susceptibility at $T_C$, together with a continuous evolution of the monoclinic angle toward the orthorhombic limit observed by X-ray diffraction.[22] However, the same authors identified strain-coupled domains and twin-related platelets, describing the transition as martensitic-like despite its thermodynamically second-order nature. Martensitic transformations usually proceed by nucleation and growth, giving rise to spatial coexistence between competing phases and strain-accommodating twinned domains. Recently, *in-situ* transmission electron microscopy has provided direct evidence that lattice defects and nanoscale confinement strongly influence the transition: dislocations locally modify the transition temperature and the evolution of twinned domains, while spatially confined regions exhibit metastable states and distinct transformation kinetics.[23] Together, these observations reinforce the martensitic-like character of the transition at the microscopic scale.

The electronic band structure of the monoclinic and orthorhombic phases has already been studied by angle-resolved photoemission spectroscopy (ARPES). Below $T_C$, the monoclinic phase exhibits a characteristic flat valence band near the Fermi level ($E_F$), which is commonly interpreted as a signature of the excitonic instability.[24,25] Above $T_C$, the electronic structure becomes semimetallic, with conduction and valence bands overlapping near the Fermi level.[26] Complementarily, scanning tunneling spectroscopy measurements have established that the electronic bandgap of monoclinic $Ta_2NiSe_5$ evolves with temperature: a well-defined bandgap (~ 0.2 eV) at cryogenic temperatures becomes partially filled with electronic states at room temperature.[27,28] Beyond thermal broadening, this evolution has been interpreted as the progressive suppression of excitonic order as temperature increases. To date, no scanning tunneling spectroscopy measurements of the orthorhombic phase or its phase transition dynamics have been reported.

Most of the current understanding of this phase transition derives from X-ray diffraction, electrical transport, optical, and ARPES measurements, which probe the global structural and electronic evolution of $Ta_2NiSe_5$ across $T_C$.[29–34] These techniques provide a momentum- and energy-resolved picture of this transition, but are inherently averaged over large sample volumes. As a result, they provide limited information about the nanoscale evolution, including the role of point and extended defects in governing its dynamics. The microscopic picture of this phase transition, including phase nucleation and propagation, as well as how defects modify these dynamics, remains largely inaccessible to these probes. Scanning tunneling microscopy and spectroscopy (STM/STS) are ideal techniques for retrieving the local electronic density of states with atomic-scale resolution. This enables the identification of spatially confined electronic variations such as charge-density-wave domains, vacancy defects, charge puddles, and step edges.[35–39] This sensitivity also makes it possible to distinguish polymorphic and polytypic phases even with minimal structural variations.[40–42]

In this work, we used variable-temperature STM/STS to probe the real-space evolution of the phase transition in $Ta_2NiSe_5$ at the nanoscale. We investigated three characteristic surface regions near $T_C$: pristine, near an isolated point-defect, and near a step edge. By tracking spectroscopic signatures upon *in-situ* heating cycles, we identify a time-dependent phase coexistence regime followed by the growth of the high-temperature orthorhombic domains. The phase transition propagates anisotropically perpendicular to the Ta-Ni-Ta chains in pristine regions, whereas point-defects and step edges act locally as nucleation centers, suppressing this intrinsic anisotropy. We ascribe these findings to a martensitic-like phase transition at the surface, where distinct monoclinic and orthorhombic phases coexist as spatially segregated domains with well-defined boundaries, in contrast to a purely second-order phase transition.

## RESULTS AND DISCUSSION

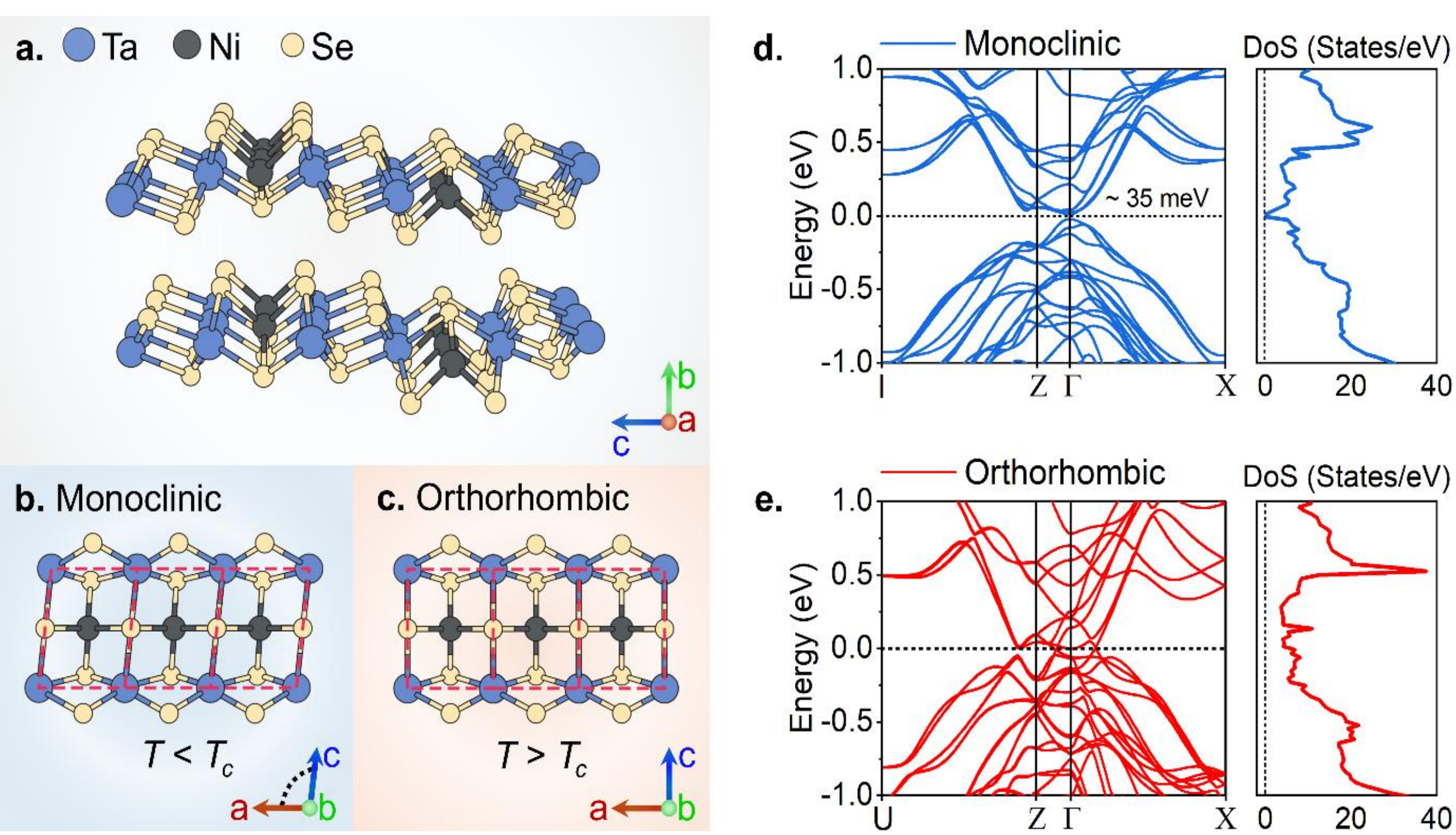

**Figure 1.** Structural and electronic structure comparison between monoclinic ($T < T_C$) and orthorhombic ($T > T_C$) $Ta_2NiSe_5$. **(a)** Side-view representation of the crystalline structure along the *b*-axis. Atomic species are color-coded as Ta (blue), Ni (gray), and Se (yellow). **(b)** Top-view comparison of the monoclinic and **(c)** the orthorhombic $Ta_2NiSe_5$ viewed along the *ac*-plane. Dashed red lines indicate the structural shear distortion associated with the phase transition, where the lattice angle changes from $\beta \approx 90.6°$ in the monoclinic phase to $\beta = 90.0°$ in the orthorhombic phase. DFT calculations (GGA-PBE) and corresponding density of states (DoS) for the **(d)** monoclinic and **(e)** orthorhombic phases are included for completeness and as a reference for the electronic structure discussion that follows.

The monoclinic (C2/c) and orthorhombic (Cmcm) phases of $Ta_2NiSe_5$ consist of corrugated Ta-Ni-Ta chains sandwiched by Se atoms via covalent bonding, with adjacent layers stacked along the *b*-axis and coupled by van der Waals forces, as shown in Figure 1a. The periodic repetition of zigzag chains leads to strong in-plane anisotropy, as previously reported in the literature.[43–47] The structural difference between the monoclinic and orthorhombic phases is a shear distortion, manifested as a deviation of the interchain angle from $\beta \approx 90.6°$ in the monoclinic phase, which continuously transitions to the higher-symmetry $\beta = 90.0°$ in the orthorhombic phase.[48] It was also observed that the β lattice-parameter changes continuously, in a manner consistent with a second-order phase transition. Figures 1b and 1c show the top-view of the monoclinic and orthorhombic phase structures, respectively. The angle between the *a*- and *c*-axes is exaggerated for better visualization of the difference between the two phases.

This structural distortion is sufficient to induce a semiconductor-to-semimetal transition between the monoclinic and orthorhombic phases, with well-characterized modifications to the electronic structure reported in the literature.[26, 49–51] For completeness, we have also carried out density functional theory (DFT) calculations within the generalized gradient approximation (GGA), using the Perdew-Burke-Ernzerhof (PBE) functional.[52,53] Figures 1d and 1e show the band structure calculations and electronic densities of states of the monoclinic and orthorhombic phases, respectively. In the monoclinic phase, the valence-band maximum lies between the Z and Γ points, while the conduction-band minimum occurs at Γ, resulting in an indirect bandgap of 35 meV. In the orthorhombic phase, the gap collapses to a small band overlap, yielding a semimetallic dispersion.

To distinguish the monoclinic and orthorhombic phases at the nanoscale, we first examined their structural and electronic characteristics in defect-free regions using variable-temperature STM/STS under identical tunneling conditions (V = 0.30 V, I = 500 pA). Figures 2a and 2b show STM topographic images of a pristine surface region at 300 K and 335 K, respectively. Visually, both topographic images appear similar, since the structural phase transition is subtle and preserves the overall zigzag arrangement of the Ta-Ni-Ta chains, with the transition instead reflected in a small shift of the inter-chain spacing. To quantify the structural change across the transition, we measured the inter-chain distance between adjacent Ta-Ni-Ta units from line profiles extracted along the *c*-direction. The resulting distributions, shown in the histograms of Figures 2c and 2d, yield average spacings of (1.59 ± 0.02) nm at 300 K and (1.69 ± 0.03) nm at 335 K, corresponding to a relative change of approximately 6%.

Representative differential conductance (dI/dV) spectra of the monoclinic (300 K) and orthorhombic (335 K) phases, each averaged over 25 curves to minimize thermal and electronic fluctuations, are shown in Figures 2e and 2f. The global minimum of the 335 K spectrum lies above that at 300 K, and the larger dI/dV values indicate an increased superposition of states crossing the Fermi energy, consistent with a more pronounced metallic character. Besides such feature, the two spectra share a similar line shape at both bias polarities, except for a softened peak near +0.40 V present only in the orthorhombic phase. We did not use this feature for phase classification, as its weak contrast and higher-bias origin make it more susceptible to tip-induced artifacts and defect-related shifts. Instead, we adopted the integrated dI/dV signal within a ±0.15 V window around $E_F$ as a classification criterion, which we show below to be robust across the different electronic environments studied here.

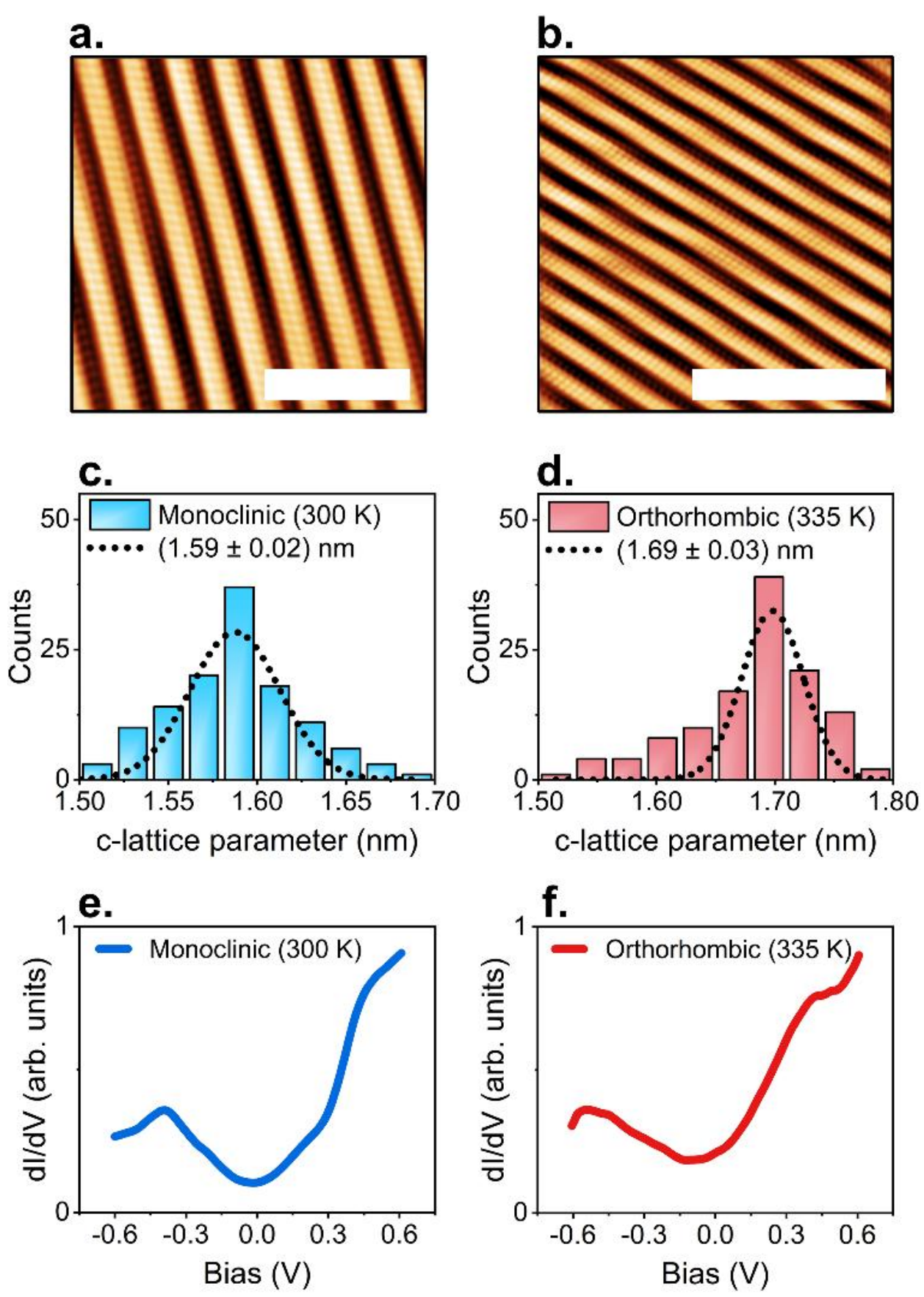

**Figure 2.** Variable-temperature STM/STS characterization of the monoclinic and orthorhombic phases. **(a)** STM image of a monoclinic region acquired at 300 K and **(b)** of an orthorhombic region acquired at 335 K. Scale bars are 5 nm and 10 nm, respectively. **(c)** Histograms of the inter-chain distance extracted from line profiles along the c-direction for the monoclinic phase, yielding an average c-lattice parameter of (1.59 ± 0.02) nm, and **(d)** the orthorhombic phase, yielding (1.69 ± 0.03) nm. **(e)** Representative dI/dV spectra, each averaged over 25 curves, for the monoclinic phase at 300 K and **(f)** for the orthorhombic phase at 335 K.

As shown in the histograms of Figures 3a and 3b, the average integrated dI/dV within a ±0.15 V window is (2.3 ± 0.1) nA for the monoclinic phase and (17.6 ± 0.3) nA for the orthorhombic phase. This nearly one-order-of-magnitude contrast between the two phases provides an unambiguous and quantitative criterion for phase identification at the nanoscale. In each case, the histogram was fit to a Gaussian distribution, and the reported values correspond to the mean and standard deviation obtained from a statistical analysis of over 1600 dI/dV curves extracted from spectroscopic maps. Using this metric, we then analyzed a series of temperature-dependent tunneling spectroscopic measurements across $T_C$ (320-335 K) to determine the onset of the phase transition.

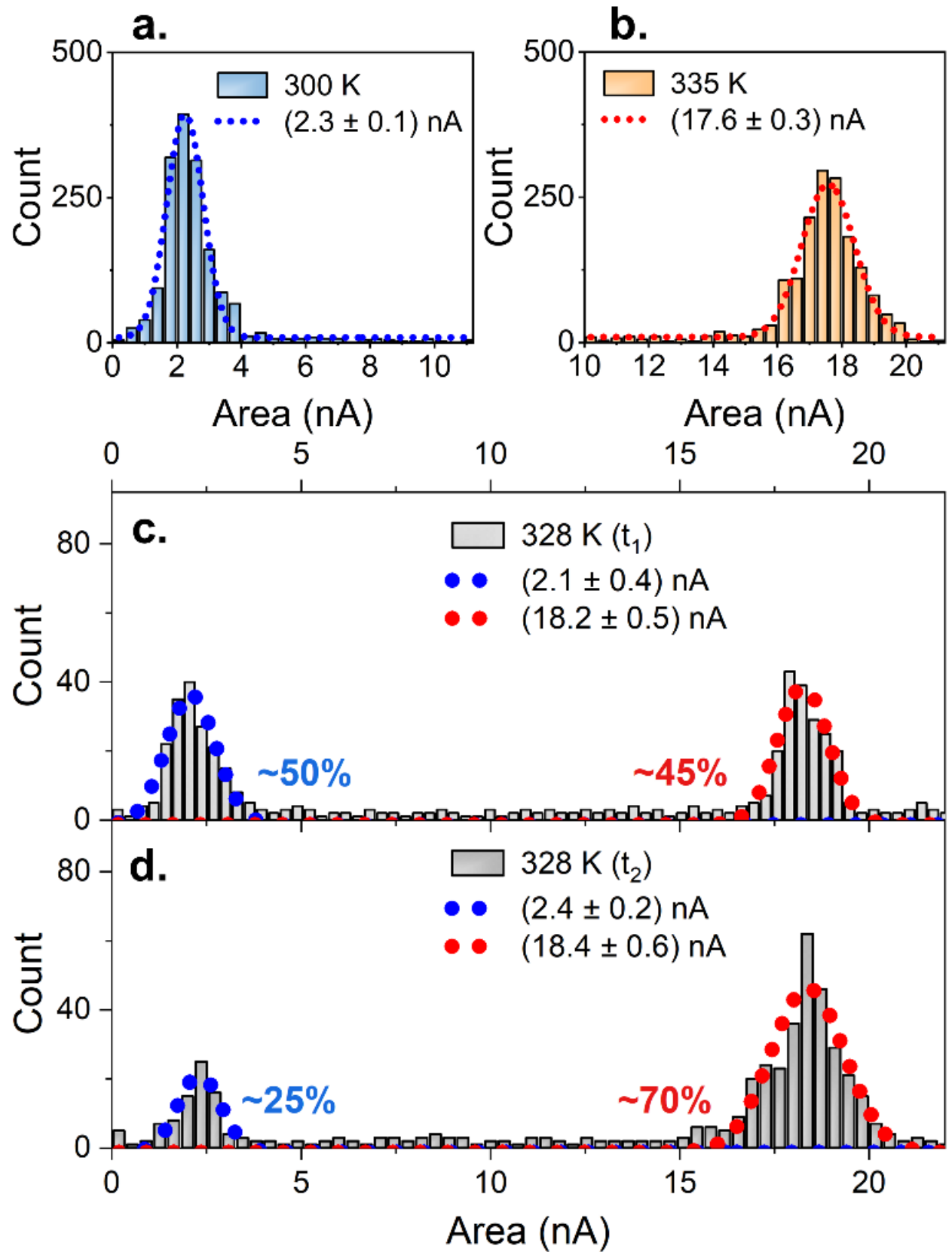

**Figure 3.** Histograms of the integrated dI/dV value over the bias window of ±0.15 V for **(a)** the monoclinic phase with an area of (2.3 ± 0.1) nA and **(b)** the orthorhombic phase with (17.6 ± 0.3) nA. The difference in integrated spectral weight serves as the phase-classification metric used throughout this work. **(c)** At 328 K, both electronic signatures are observed simultaneously, corroborating a regime of phase coexistence. The bimodal distributions are (2.1 ± 0.4) nA and (18.2 ± 0.5) nA for monoclinic and orthorhombic signatures, respectively. **(d)** Histogram acquired 40 minutes later at the same temperature and region, illustrating the temporal evolution of the phase coexistence. The integrated values are (2.4 ± 0.2) nA and (18.4 ± 0.6) nA for monoclinic and orthorhombic phases, respectively.

At 328 K, we observed the coexistence of monoclinic and orthorhombic spectroscopic signatures, seen in the bimodal histogram of Figure 3c. A double-Gaussian fit yields peak positions of (2.1 ± 0.4) nA and (18.2 ± 0.5) nA, consistent with the values obtained for the pure monoclinic and orthorhombic phases. A second histogram over the same region was obtained 40 min later, as shown in Figure 3d. The distribution remains bimodal, with peak positions of (2.4 ± 0.2) nA and (18.4 ± 0.6), but the relative weight of the two populations changes significantly, with the orthorhombic fraction increasing from approximately 45% to 70%. This indicates that the transition dynamics that take place at the surface at $T_C$ do not occur abruptly or homogeneously throughout the material, but instead proceed via a finite-time nucleation and growth process. We next examine the spatial arrangement and the temporal evolution of this phase transition.

Spatially-resolved maps of the integrated dI/dV signal and the corresponding topography for pristine, point-defect, and step edge regions are presented in Figures 4a, 4b, and 4c. These measurements were conducted right after thermal stabilization at 328 K. The spectroscopic maps correspond to three energy windows: I = [−0.50, −0.30] V, II = [−0.15, +0.15] V, and III = [+0.30, +0.50] V. In each panel, the images are arranged from top to bottom as windows I, II, III, and the topography. Figure 4d, 4e, and 4f show representative dI/dV spectra of the monoclinic (blue) and orthorhombic (red) phases at each structural environment. The use of bias-integrated rather than single-bias maps is justified by the thermal broadening inherent to STS measurements. Since the intrinsic energy resolution of STS is limited by the derivative of the Fermi-Dirac distribution, with a full width at half maximum of approximately 3.5 $k_BT$, this corresponds to an effective broadening of ~90-100 meV over this temperature range.

In the pristine region, shown in Figure 4a, window I (negative bias) presents only weak contrast between the two electronic regimes, consistent with the similar intensity and lineshape of the monoclinic and orthorhombic tunneling spectra at negative bias shown in Figure 3d. In window II, the higher-intensity domains (orthorhombic phase) nucleate preferentially along the Ta-Ni-Ta chain direction, matching the topography shown in the bottom panel. One notices that the largest intensity variations are observed along the in-plane direction perpendicular to the surface corrugation chains. Therefore, this behavior reveals an anisotropic distribution of the local density of states (LDOS). Window III (positive bias) also presents the same spatial pattern with strong contrast. The peak near +0.40 V seen in the orthorhombic spectrum, which falls within the [+0.30, +0.50] V integration range, accounts for the observed contrast despite the comparable overall intensity of the two spectra at positive bias.

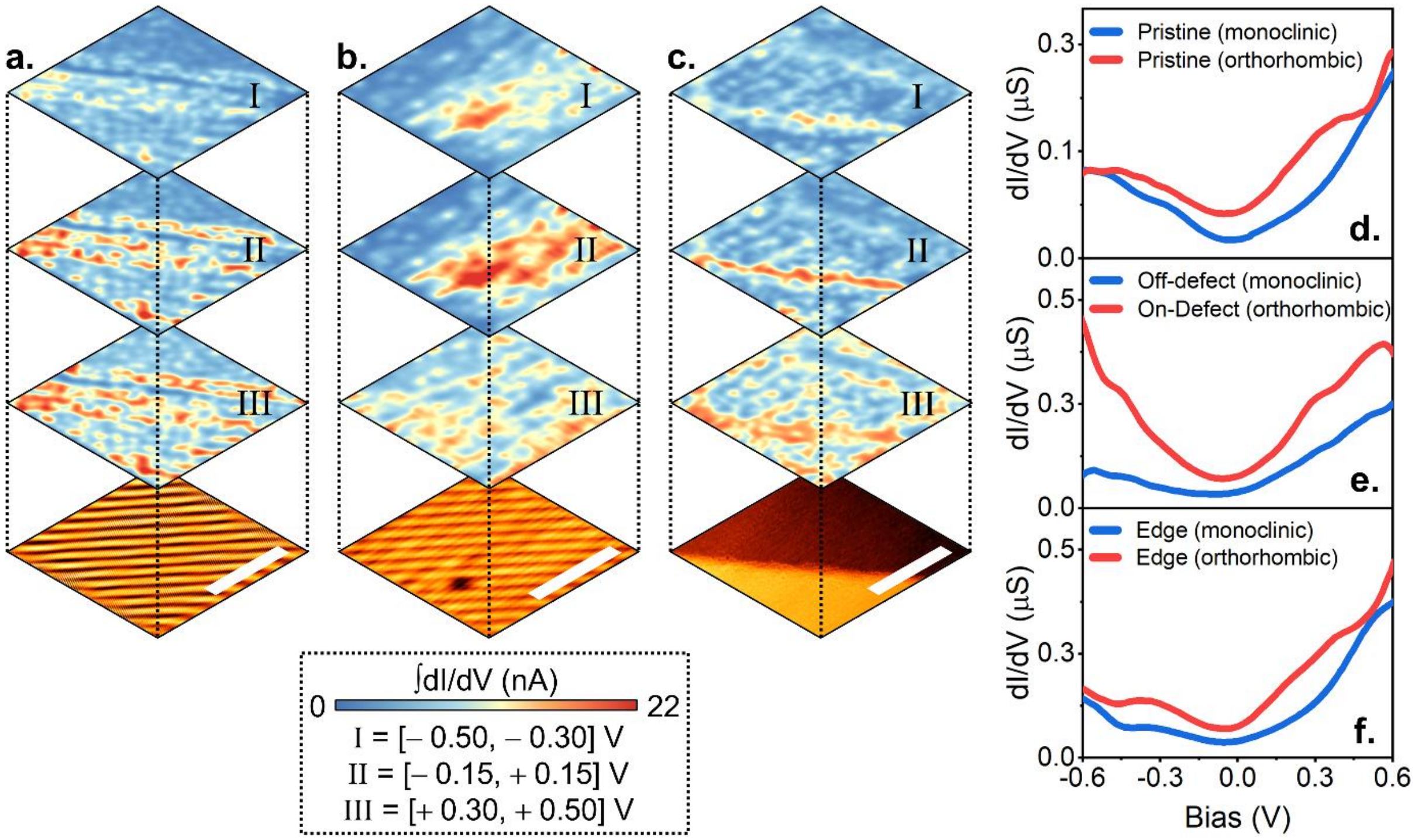

**Figure 4.** STM and STS characterization of pristine, point-defect, and step edge regions immediately after reaching 328 K. Integrated dI/dV maps are shown for three bias windows, I = [−0.50, −0.30] V, II = [−0.15, +0.15] V, and III = [+0.30, +0.50] V, along with the corresponding topographic images (lower panel) for **(a)** pristine, **(b)** point-defect, and **(c)** step edge regions. Scale bars are 10 nm, 10 nm, and 20 nm, respectively. The colorbar represents the integrated dI/dV signal (0-22 nA). Representative dI/dV spectra averaged over 25 curves acquired from monoclinic (blue) and orthorhombic (red) regions are shown for the **(d)** pristine surface, **(e)** off-defect (monoclinic) and on-defect (orthorhombic) regions, and **(f)** step edge.

In contrast to the preferential propagation perpendicular to the Ta-Ni-Ta chains observed in pristine regions, orthorhombic phase nucleation is locally modified near point-defects, as shown in the spectroscopic and topographic data of Figure 4b. In window I (negative bias), the contrast near the defect is enhanced relative to pristine areas, as clearly resolved in the tunneling spectra of monoclinic (blue) and orthorhombic (red) shown in Figure 4e. We observe that the orthorhombic phase exhibits a modified electronic structure near the defect, with pronounced LDOS at negative bias. This is consistent with previous reports of top-surface defects in $Ta_2NiSe_5$, which observed a similar dI/dV enhancement at negative bias in the low-temperature monoclinic phase.[38] Here, we extend this observation to defect-induced electronic variations in the orthorhombic phase. In window II, the region surrounding the defect continues to exhibit a high dI/dV signal, indicating phase nucleation originating from the defect. However, in window III, the observed integrated LDOS values are relatively more homogeneous with respect to Windows I and II. We ascribe this result for window III as follows: in the vicinity of the point-defect, additional degrees of freedom for lattice relaxation exist. In such conditions, the surrounding atoms may also experience conditions in which structural and electronic states that cannot be achieved in the pristine surfaces. Under such locally relaxed conditions, the surrounding atoms may access structural and electronic configurations that are not accessible in the pristine surface, which could account for the more homogeneous LDOS observed in window III near the defect.

Near step edges with heights larger than the distance between neighboring Van der Waals gaps, the orthorhombic phase also nucleates preferentially along the edge, as observed in Figure 4c. The step analyzed here has a height of 3.70 nm. Window I (negative bias) shows only a subtle contrast between the two phases, similar to the pristine case. The tunneling spectra of both phases at the step edge, shown in Figure 4f, are similar in the negative bias range. In window II, however, we observe a pronounced variation in the integrated dI/dV signal, with the orthorhombic phase emerging locally at the step edge. In window III (positive bias), the softened peak at approximately +0.40 V again produces contrast, although some monoclinic-associated regions exhibit a higher integrated dI/dV signal. Similarly to the point-defect regions, the presence of a large topographic step introduces a distinct aperiodic boundary condition for lattice relaxation along the in-plane direction, enhancing or suppressing electronic states that would be found in pristine regions. Therefore, the nucleation of the orthorhombic phase starts preferentially near the point-defect and step edge. In particular, unlike the point-defect region, the step edge does not exhibit the distinct enhancement of the LDOS at negative bias observed at the defect.

As demonstrated above, phase coexistence emerges at 328 K, enabling a real-space visualization of the phase transition during heating cycles from the monoclinic to the orthorhombic phase. Figures 5a, 5b, 5c, 5d, 5e, 5f, 5g, 5h, and 5i present sequential spatially-resolved dI/dV maps acquired at 328 K over pristine, point-defect, and step edge regions. Consecutive maps were recorded at fixed sample positions with a constant acquisition interval of 40 min, corresponding to the time required to complete a full topographic and spectroscopic map. To facilitate visualization of the phase evolution, each integrated dI/dV map is accompanied by a supplementary phase map obtained using the classification criterion previously established. For our measurement conditions, pixels with integrated dI/dV values between 0 and 5 nA are assigned to the monoclinic phase, values between 16 and 22 nA to the orthorhombic phase, while intermediate values between 6 and 15 nA are classified as a regime where the local electronic response does not allow an unambiguous phase assignment. These regions are represented in blue, red, and gray, respectively. Throughout the rest of the discussion, phase fractions are reported as (monoclinic/orthorhombic/intermediate) percentages.

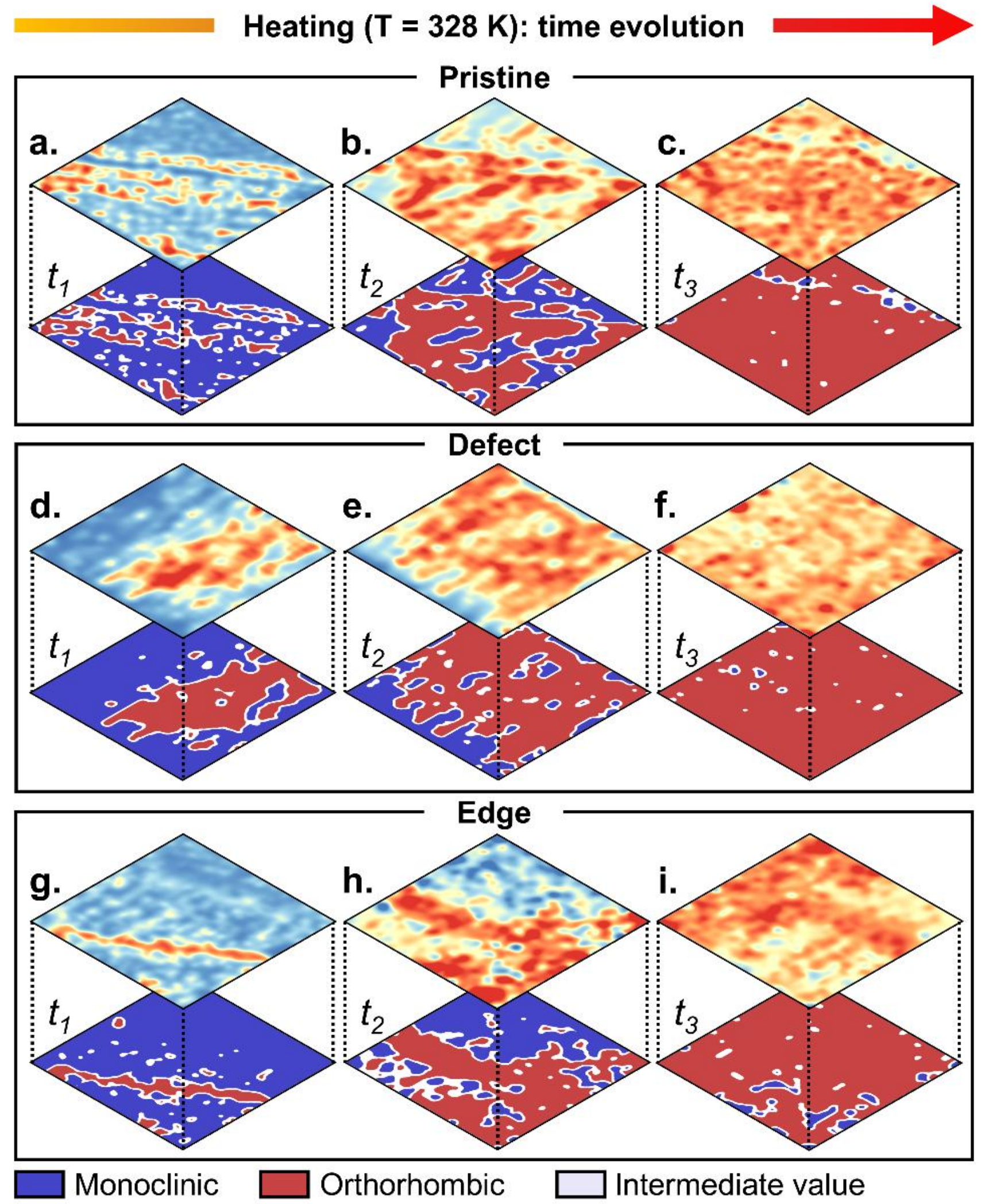

**Figure 5.** Time-resolved nanoscale evolution of the monoclinic-to-orthorhombic phase transition in $Ta_2NiSe_5$ at 328 K using STM/STS, resolved across pristine, point-defects, and step edges regions. Top panels: integrated dI/dV maps (±0.15 V bias window). Bottom panels: corresponding phase maps (blue: monoclinic; red: orthorhombic; white: intermediate value). Consecutive maps within each row were acquired at 40 min intervals. **(a)** Pristine terrace, $t_1$. **(b)** Pristine terrace, $t_2$. **(c)** Pristine terrace, $t_3$. **(d)** Point-defect, $t_1$. **(e)** Point-defect, $t_2$. **(f)** Point-defect, $t_3$. **(g)** Step edge, $t_1$. **(h)** Step edge, $t_2$. **(i)** Step edge, $t_3$.

The temporal evolution of the phase transition in the pristine region is shown in Figures 5a, 5b, and 5c. At the initial stage (Figure 5a), orthorhombic domains nucleate preferentially along the Ta-Ni-Ta chains, with phase fractions of (75/20/5)%. After one acquisition interval (Figure 5b), the initially nucleated domains expand laterally while smaller orthorhombic nuclei continue to grow throughout the field of view, resulting in phase fractions of (38/52/10)%. This evolution is consistent with growth dominated by the advance of existing phase boundaries rather than by continuous nucleation of new domains. By the final stage (Figure 5c), the orthorhombic phase occupies nearly the entire scanned region, leaving only isolated monoclinic islands, with phase fractions of (3/92/5)%.

A different behavior is observed for the point-defect region, shown in Figures 5d, 5e, and 5f. In the first stage

(Figure 5d), the orthorhombic phase nucleates preferentially around the defect, producing a substantially larger transformed area than in the pristine region after the same time interval. The phase percentages retrieved are (62/30/8)%. This accelerated transition indicates that the point-defect locally lowers the nucleation barrier, promoting earlier stabilization of the orthorhombic structure. As the transition progresses (Figure 5e), the orthorhombic domain continues to expand away from the defect center, with phase percentages of (24/70/6)%. By the final frame (Figure 5f), the orthorhombic phase dominates nearly the entire field of view, with only small residual monoclinic regions remaining (3/95/2)%. Compared with the pristine terrace, the transition near point-defects proceeds faster and from a localized nucleation center.

Finally, Figures 5g, 5h, and 5i show the transition dynamics near a step edge. The orthorhombic phase initially develops along the step (Figure 5g), with domains largely confined to its vicinity and only a limited number extending into the terrace, shown in Figure 5g, with phase fractions of (83/12/5)%. In the subsequent frame (Figure 5h), a small lateral drift associated with thermal broadening is observed. However, the overall evolution is unambiguous, with the orthorhombic phase continuing to expand outward from the step edge into the surrounding terrace. The corresponding phase fractions are (40/46/14)%. By the final stage (Figure 5i), the orthorhombic phase dominates most of the scanned area, with phase fractions of (8/89/3)%. Notably, the orthorhombic phase nucleates in a well-localized region right at the step edge and subsequently spreads preferentially into the upper terrace, while the lower terrace shows comparatively little orthorhombic growth throughout the observed time window. This asymmetry likely explains why the step edge region shows the slowest net conversion of the three: nucleation on the lower terrace appears kinetically hindered, so the transformed phase remains concentrated above the step for an extended period before the lower terrace begins to convert.

We compared quantitatively the phase transition dynamics across the three regions, and Figures 6a, 6b, 6c, 6d and 6e summarize the time-dependent evolution of the monoclinic and orthorhombic fractions extracted from the previous analysis. Although all three regions converge to near-complete transformation by $t_3$, the growth evolution differs among them. For instance, the pristine region exhibits intermediate and approximately monotonic kinetics of the orthorhombic phase transition throughout the observed interval (+32% and +40%), shown in Figure 6a. At the point-defect, shown in Figure 6b, the orthorhombic fraction already starts from a higher value at $t_1$ (30%) and shows the largest increment between $t_1$ and $t_2$ (+40%), before decelerating in the final interval as the phase transition approaches saturation. This behavior is consistent with the defect acting as a heterogeneous nucleation site that locally lowers the energy barrier, promoting early formation of the orthorhombic phase. The subsequent slowdown simply reflects the shrinking monoclinic area available for conversion.

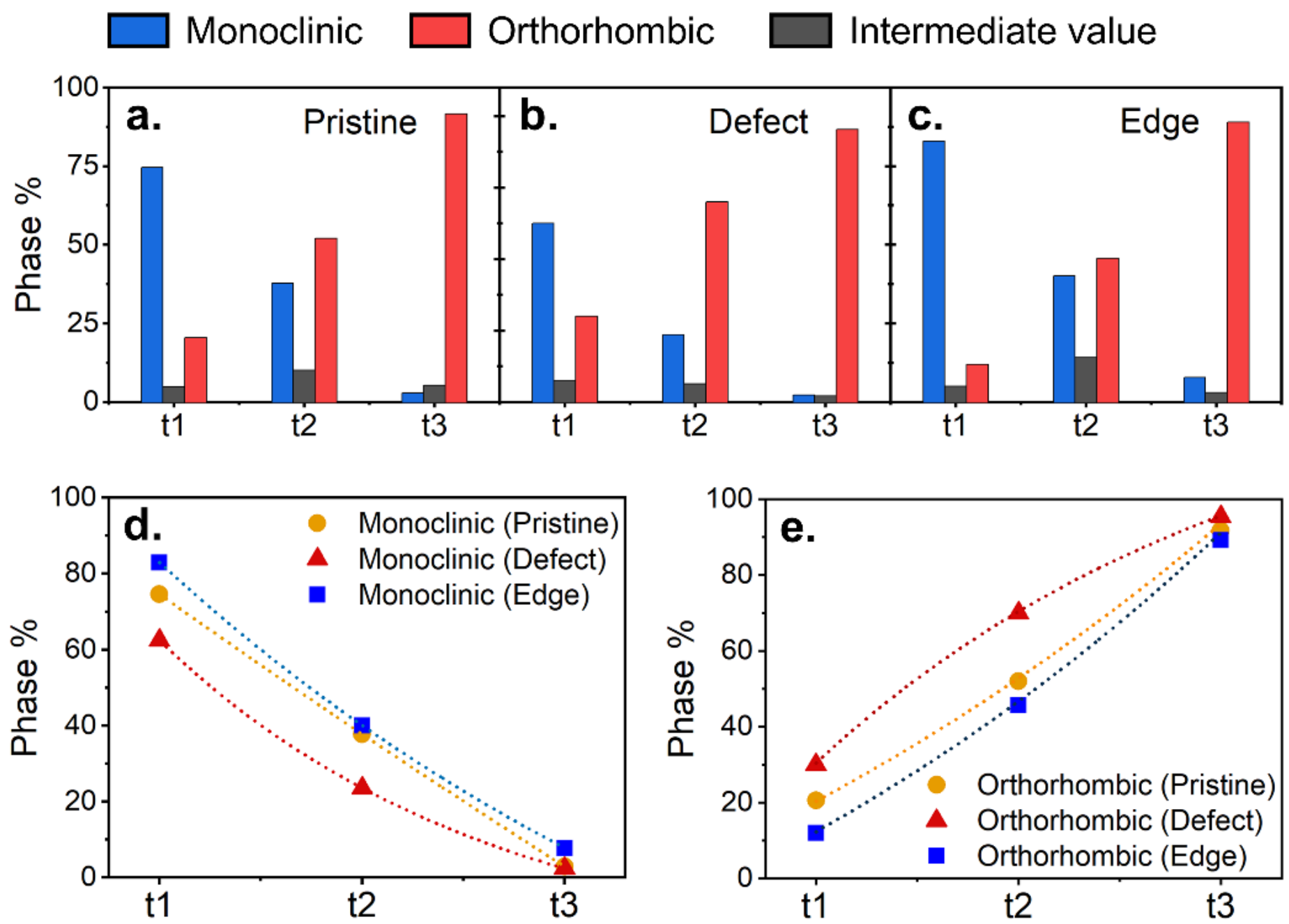

**Figure 6.** Quantitative comparison of phase transition kinetics across pristine, point-defect, and step edge regions. Bar charts showing the temporal evolution of the monoclinic (blue), orthorhombic (red), and intermediate (gray) phase fractions for **(a)** the pristine, **(b)** point-defect, and **(c)** step edge regions at $t_1$, $t_2$, and $t_3$. **(d)** Comparison of the monoclinic phase fraction decay across the three regions. The point-defect region exhibits the fastest decay, while the pristine and step edge regions display comparable kinetics. **(e)** Comparison of the orthorhombic phase fraction growth across the three regions, showing the fastest growth at the point-defect, followed by comparable rates for the pristine and step edge regions.

The step edge region displays the opposite behavior, evidenced in Figure 6c: the initial orthorhombic fraction is the lowest among the three regions (12.0%), yet the largest kinetic gain occurs in the intermediate interval, +34% between $t_1$-$t_2$ and +43% between $t_2$-$t_3$. Step edges lower the local nucleation barrier, providing energetically favorable sites for the initial emergence of the orthorhombic phase, but the same structural discontinuity may subsequently hinder the advance of the phase boundary, for instance through local strain fields or broken lattice continuity that pin the interface, limiting the net conversion despite a persistently positive growth rate.

Figures 6d and 6e summarize these trends directly by overlaying the monoclinic decay and orthorhombic growth curves for the three regions. The point-defect region consistently shows the steepest monoclinic decay and orthorhombic growth, confirming its role as the most efficient nucleation site. In contrast, the pristine and step edge regions display comparable overall kinetics despite their very different early-stage behavior, indicating that once nucleation has occurred, the subsequent domain-boundary propagation proceeds at a similar rate regardless of the initial nucleation geometry. This comparison reinforces the picture that structural heterogeneities primarily influence the onset of the transition by lowering the local nucleation barrier rather than the intrinsic velocity of the phase boundary once formed.

It is worth noting that this time evolution of the phase coexistence could, in principle, be associated with temperature stabilization of the surface in the measurement regions. However, two observations argue against this interpretation. First, the phase coexistence was confined to areas no larger than $20 \times 20$ nm$^2$, over which temperature gradients and local thermal fluctuations are not expected to be significant. Second, throughout the measurements, the temperature sensor near the sample recorded fluctuations no larger than $\pm 0.2$ K, ruling out macroscopic thermal instability as the origin of the observed phase dynamics. Therefore, this spatially inhomogeneous coexistence of monoclinic and orthorhombic phases near $T_C$, with well-defined domain boundaries evolving in time at fixed temperature, is naturally interpreted as a real phase transition mechanism inherent to $Ta_2NiSe_5$.

As the bulk phase transition in $Ta_2NiSe_5$ is generally reported as second-order, the nucleation and growth dynamics observed here, with distinct phase domains separated by sharp boundaries rather than a spatially uniform order parameter evolving continuously, are inconsistent with a purely second-order scenario. Therefore, our STM/STS observations of coexisting domains with time-dependent phase fractions are consistent with the second-order-like behavior reported from bulk diffraction: the two pictures correspond to the same martensitic-like transformation viewed at different length scales, spatially averaged over many domains in the bulk case and resolved at the level of individual domains and in our real-space measurements. We therefore favor describing the surface transition as martensitic-like. We further note that surface-specific factors – local disorder, defects, and reduced dimensionality – likely renormalize the nucleation kinetics of this martensitic-like transition relative to the bulk, for instance by lowering local nucleation barriers or modifying domain-wall energetics.

## CONCLUSION

In summary, we spatially resolved the monoclinic-to-orthorhombic phase transition of $Ta_2NiSe_5$ at the nanoscale using variable-temperature STM/STS. We found that the transition proceeds through the time-dependent coexistence of monoclinic and orthorhombic domains separated by well-defined boundaries, rather than through a spatially homogeneous evolution of the structural and electronic properties across the surface. In pristine regions, the orthorhombic phase nucleates anisotropically, with the transformation preferentially propagating perpendicular to the Ta-Ni-Ta chains. In point-defects regions, top-surface vacancies act as local nucleation centers that accelerate the transition and locally suppress this anisotropy, resulting in the fastest overall conversion to the orthorhombic phase. Step edge regions, in contrast, exhibit a pronounced asymmetry between the upper and lower terraces in their transition dynamics, indicating that the step acts as a kinetic barrier to phase conversion and leads to a slower overall transformation than in both pristine and point-defect regions. Taken together, these observations are best captured by a martensitic-like description of the surface transition, characterized by phase coexistence, path-dependent domain evolution, and defect-sensitive nucleation. Our results establish STM/STS as a powerful approach for resolving the nanoscale heterogeneity of structural phase transitions in correlated quantum materials and highlight the central role of surface defects and boundary conditions in renormalizing transition dynamics relative to those inferred from bulk-sensitive probes.

## METHODS

**Sample Preparation.** Bulk single crystals of $Ta_2NiSe_5$, synthesized by the flux-zone growth method with a certified purity of 99.9999%, were purchased from 2D Semiconductors Inc. The crystals were cleaved immediately before insertion into the scanning tunneling microscope chamber to ensure a clean surface and minimize contamination prior to measurements.

**Scanning Tunneling Microscopy and Spectroscopy.** STM and STS measurements were performed using an Omicron variable-temperature STM operating under ultrahigh vacuum regime, with a base pressure better than $3 \times 10^{-10}$ mbar. A single tungsten (W) tip was used throughout the entire work to ensure consistency among measurements. The tip was prepared by standard electrochemical etching in a NaOH solution and subsequently conditioned *in-situ* by electron-beam heating at 1000 V.

All topographic and dI/dV maps were acquired with a sample bias of 0.30 V and a tunneling current setpoint of 500 pA to eliminate possible artifacts arising from electrostatic effects. Controlled heating of the sample was achieved using a Lake Shore 332 temperature controller, which provided stable temperature regulation over the investigated range. Based on the controller specifications, the temperature uncertainty above room temperature is ± (1 K + 0.5% × T).

**Density Functional Theory.** The electronic band structure and electronic density of state calculations presented in this work were carried out within the density functional theory (DFT) framework. We used an implementation of the generalized gradient approximation (GGA) with Perdew-Burke-Ernzerhof (PBE) on the QuantumESPRESSO package. The projector augmented wave (PAW) technique was incorporated to describe the interactions between valence electrons and ions. Structural optimization was performed until forces on the atoms were below the threshold of 0.01 eV/nm using a cutoff of 520 eV for the plane wave basis set.

## AUTHOR INFORMATION

### Corresponding Author

**Rogério Magalhães-Paniago -** Department of Physics, Federal University of Minas Gerais, Belo Horizonte, Minas Gerais, 30123-970, Brazil;

https://orcid.org/0000-0002-5203-0944

E-mail: rogerio.paniago0@gmail.com

### Authors

**Guilherme Rodrigues-Fontenele –** Physics Department, Federal University of Minas Gerais, Belo Horizonte, Minas Gerais, 30123-970, Brazil;

https://orcid.org/0000-0002-3650-1101

**Gabriel Fontenele -** Physics Department, Federal University of Minas Gerais, Belo Horizonte, Minas Gerais, 30123-970, Brazil;

https://orcid.org/0000-0001-9191-8746

**Angelo Malachias** – Physics Department, Federal University of Minas Gerais, Belo Horizonte, Minas Gerais, 30123-970, Brazil;

https://orcid.org/0000-0002-8703-4283

## ACKNOWLEDGMENT

The authors acknowledge financial support from the Brazilian funding agencies CNPq, CAPES, and FAPEMIG.